%% file: disk_chemistry_ws41.tex
\documentclass[10pt,twoside,russian]{winter}
\RequirePackage[utf8x]{inputenc}
\RequirePackage[T2A]{fontenc}
\RequirePackage[russian,english]{babel} \RequirePackage{epsfig}
\RequirePackage{amsmath} \RequirePackage{amssymb}
\RequirePackage{amsfonts} \RequirePackage{longtable}
\usepackage{chapterbib}
\usepackage{indentfirst}
\usepackage{cite}

\usepackage{lscape}

\begingroup
\catcode`\+\active\gdef+{\mathchar8235\nobreak\discretionary{}%
 {\usefont{OT1}{cmr}{m}{n}\char43}{}}
\catcode`\-\active\gdef-{\mathchar8704\nobreak\discretionary{}%
 {\usefont{OMS}{cmsy}{m}{n}\char0}{}}
\catcode`\=\active\gdef={\mathchar12349\nobreak\discretionary{}%
 {\usefont{OT1}{cmr}{m}{n}\char61}{}}
\endgroup

\AtBeginDocument{\mathcode`\==32768 \mathcode`\+=32768
\mathcode`\-=32768 }

\def\eqalign#1{\null\,\vcenter{\openup\jot\m@th
  \ialign{\strut\hfil$\displaystyle{##}$&$\displaystyle{{}##}$\hfil
      \crcr#1\crcr}}\,}
\def\eqalignleft#1{\null\,\vcenter{\openup\jot\m@th
  \ialign{\strut$\displaystyle{##}$\hfil&$\displaystyle{{}##}$\hfil
      \crcr#1\crcr}}\,}

\def\b#1:{{\bf#1}, }

\def\beq#1{\begin{equation}\label{#1}}
\def\eeq{\end{equation}}

\def\b{\begin{eqnarray}}
\def\earr{\end{eqnarray}}

\def\fun#1#2{\lower3.6pt\vbox{\baselineskip0pt\lineskip.9pt
\ialign{$\mathsurround=0pt#1\hfil##\hfil$\crcr#2\crcr\sim\crcr}}}

\def\b{\begin{eqnarray}}
\def\earr{\end{eqnarray}}

\def\fun#1#2{\lower3.6pt\vbox{\baselineskip0pt\lineskip.9pt
\ialign{$\mathsurround=0pt#1\hfil##\hfil$\crcr#2\crcr\sim\crcr}}}

\makeatletter
\renewcommand{\footnoterule}{\kern-3\p@
 \hrule width .4\columnwidth
 \kern 2.6\p@}
\renewcommand{\@makefnmark}{}
\renewcommand{\@makefntext}[1]{\hspace{2em}\hbox{\hss#1}}

\addto\captionsrussian{%
\def\abstractname{}
}

\renewcommand{\@biblabel}[1]{#1.}
\makeatother

\newcommand{\pud}{\hbox to 0.7em {\hspace{0.2em}.\hss $^d$}}
\newcommand{\pus}{\hbox to 0.4em {\hspace{0.01em}$''$\hss .}}
\newcommand{\pum}{\hbox to 0.7em {\hspace{0.2em}.\hss $^m$}}
\newcommand{\pug}{\hbox to 0.4em {\hspace{0.05em}.\hss $^{\circ}$}}
\newcommand{\Rim}[1]{\uppercase\expandafter{\romannumeral#1}}

\newcommand{\bnl}[1]{\begin{equation}\label{#1}}
\newcommand{\ed}{\end{equation}}

\begin{document}

\selectlanguage{russian}

\input ds_chapter.tex

\end{document}

%% file: ds_chapter.tex
\par
\hspace{0.3\textwidth}
\parbox[t]{0.65\textwidth}
{
\raggedleft
{\bf Д.~А.~Семенов$^{1}$}\\
$^{1}$Институт астрономии общества Макса Планка, Гейдельберг, Германия\\
\raggedright
}

\footnote{\copyright\ Д.~А.~Семенов, 2012}
\par
\medskip
\begin{center}
\bf
ХИМИЯ В ПРОТОПЛАНЕТНЫХ ДИСКАХ
\end{center}

\addcontentsline{toc}{subsection}{{\bf Семенов Д.~А.}Химия в протопланетных дисках}
\par
\medskip

\begin{minipage}{0.9\textwidth}
{\small Краткая аннотация.}\end{minipage} \par \vspace{2mm}
В данной обзорной лекции обсуждается текущее понимание того, как происходит химическая эволюция в протопланетных дисках --
аналогах нашей Солнечной системы, когда ей было всего несколько миллионов лет. На этой стадии эволюции, когда в таких
системах начинают образовываться планеты, сильные градиенты плотности, температуры, и интенсивности ионизирующего излучения
приводят к образованию ``слоистой'' химической структуры. В горячей, разреженной и сильно ионизованной атмосфере
присутствуют только лишь атомы, атомарные ионы, и простые фотостабильные радикалы, чья химическая эволюция определяется
ограниченным набором газофазных реакций. В более глубоких слоях диска рентгеновское и ультрафиолетовое межзвездное и звездное
излучение начинает ослабляться за счет поглощения пылью и газом, температура понижается, позволяя образовываться множеству сложных
соединений посредством газопылевой химии. Наконец, в темной, холодной и плотной экваториальной части диска
большинство молекул успевает вымерзнуть за время жизни диска, образуя сложные ледяные мантии на пылинках, на которых происходит
образование сложных (органических) молекул за счет поверхностных реакций.
\selectlanguage{english}

\begin{minipage}{0.9\textwidth}
{\small Annotation.}
In this lecture I discuss recent progress in the understanding of the chemical evolution of protoplanetary disks that resemble
our Solar system during the first ten million years. At the verge of planet formation,
strong variations of temperature, density, and radiation intensities in these disks
lead to a layered chemical structure. In hot, dilute and heavily irradiated atmosphere only simple radicals, atoms, and atomic
ions can survive, formed and destroyed by gas-phase processes. Beneath the atmosphere a partly UV-shielded, warm molecular
layer is located, where high-energy radiation drives rich chemistry, both in the gas phase and on dust surfaces. In a cold,
dense, dark disk midplane many molecules are frozen out, forming thick icy mantles where surface chemistry is active and where
complex (organic) species are synthesized.
\end{minipage}

\selectlanguage{russian}
\section{Введение}
Протопланетные диски обнаружены вокруг множества молодых звезд в Галактике (\cite{Lissauer_87a}).
Это относительно небольшие ($\sim 10-1\,000$~а.е.), мало-массивные ($\lesssim 10\%$ массы Солнца) и
коротко-живущие ($\lesssim 1-10$ миллионов лет) вращающиеся объекты, состоящие преимущественно из газа
и небольшого ($\sim 1\%$) количества пыли.
Их химический состав, а также физические условия обуславливают скорость и эффективность
образования планет, которое зависит от температуры и распределения плотности газа, непрозрачности среды, и наличия турбулентности.
Более того, некоторые молекулы и пылевые частицы являются важными
для нагрева и охлаждения газа (H$_2$, СО, OH, полиароматические углеводороды), в то время как маленькая ($\lesssim 1-10$~мкм)
пыль определяет степень непрозрачности диска для излучения.

За последнее десятилетие наши знания о протопланетных дисках, в особенности, об их химическом составе,
значительно продвинулись вперед. Молекулярный водород -- главный компонент газа, будучи симметричной
молекулой без постоянного дипольного момента, не наблюдаем в дисках при температурах ниже около 500~K, 
поэтому вместо H$_2$ приходится использовать другие, гораздо менее распространенные молекулы (см. таблицу~\ref{obs_methods}).
Их вращательные эмиссионные переходы наблюдаются, обычно, в миллиметровом диапазоне длин волн, а колебательные или
колебательно-вращательные -- в инфракрасном. Преимущество наблюдений в (суб-)миллиметровом диапазоне это наличие огромного числа
молекулярных линий, а также относительная прозрачность земной атмосферы. 

Главная трудность таких наблюдений это относительно малый угловой размер
даже самых больших и близких протопланетных дисков ($\lesssim 10^{''}$ для диска вокруг ДМ~Тельца), что требует
радио-интерферометрических наблюдений для получения карт распределения излучения в линиях молекул по объекту, например, с помощью 
интерферометра Плато де Бюр (Plateau de Bure interferometer) во Франции и Субмиллиметрового Интерферометра
(Submillimeter Array) в США. Зачастую, чтобы оценить общее количество молекул и интенсивность излучения их линий в каком-либо
диске, используют наблюдения низкого разрешения на крупных радиотелескопах в широком ($>1-8$~ГГц) диапазоне частот, например,
30-м антенне IRAM (ESO, Испания),12-м антенне APEX (ESO, Чиле), и 15-м антенне телескопа Джеймса Клерка Максвелла (James
Clerk Maxwell Telescope, Гавайи).

Другая трудность состоит в том, что анализ эмиссионных молекулярных линий, к сожалению, требует построения априорной модели
о распределениях температуры, плотности, и химических концентраций по диску, а также моделирования переноса излучения
(зачастую используется приближение локального термодинамического равновесия, ЛТР). В случае необходимости точного моделирования
переноса в линиях
для какой-то выбранной молекулы надо еще где-то найти данные о столкновительных сечениях для разных переходов, например,
в Leiden Atomic and Molecular DAtabase (LAMDA; \cite{lamda}). Таким образом, цикл определения различных параметров
дисков требует реалистичную модель объекта, а также его (интерферометрических) наблюдений в линиях диагностических молекул, да еще
и в разных переходах, которые возбуждаются при определенном диапазоне температур и плотностей. Используя такой трудоемкий
(обычно, итеративный) подход удается оценить такие важные параметры дисков как радиальные и вертикальные распределения температуры
и плотности газа, а также реконструировать поле скоростей и лучевую концентрацию излучающих молекул и степень их вымерзания в
холодных частях диска
(см. обзоры \cite{Bergin_ea07} и \cite{Semenov_ea10a}). До недавнего времени таким способом можно было изучать
внешние части дисков, $\gtrsim 50-200$~а.е. С помощью космических (``Спитцер'', ``Гершель'') и наземных (Кек, VLT, Субару)
телескопов молекулы впервые были обнаружены во внутренних частях протопланетных дисков, $1-50$~а.е., где в последующем будут
образовываться (либо уже образуются) планетные системы (см., например, \cite{Lahuis_ea06,Salyk_ea08}).

Протопланетные диски обычно частично непрозрачны для фотонов с длиной волны меньше, чем $\sim 100$~мм.
Поэтому используют наблюдения теплового излучения пыли в континууме на миллиметровых длинах волн, чтобы оценить
массу дисков (с большой погрешностью из-за неизвестного соотношения массы пыли к массе
газа, и величины непрозрачности). Также удается определить степень роста пылевых частиц посредством сравнения интенсивности
теплового излучения пыли на инфракрасных, миллиметровых и сантиметровых длинах волн -- если рост есть, то падение интенсивности
с длиной волны происходит медленнее. Таким образом было обнаружено, что во многих дисках существуют, как минимум, ``пылинки''
миллиметрового и сантиметрового размера (\cite{Cortes_ea09}).

Наблюдения в инфракрасном диапазоне с поверхности Земли сильно ограничены атмосферными окнами
прозрачности, а потому требуют космических телескопов. Однако, посредством инфракрасной спектроскопии удается узнать многое о
минералогическом составе пыли в дисках. Так, было найдено, что в составе пыли в дисках присутствуют аморфные углеродистые
соединения (неизвестной природы), аморфные и кристаллические силикаты (например, MgSiO$_3$ and Mg$_2$SiO$_4$), и сложные
молекулярные льды, включающие органические вещества (например, H$_2$O, CO, CO$_2$, HCN, CH$_3$OH). Изучая форму силикатных
эмиссионных полос на 10 и 20~мкм удалось показать, что в дисках присутствуют силикаты с различными соотношением Mg и Fe,
размерами от субмикронных до нескольких микрон, а также различной степенью кристалличности, которая может еще и изменяться по
диску (\cite{Bouwman_ea08}).

К счастью, дополнительная информация об условиях, в которых образовываются планеты, может быть получена путем детального
анализа химического, минералогического, и петрологического состава разнообразных образцов материалов, из которых состоят
метеориты и частицы кометной/зодиакальной пыли (\cite{Bradley_05}). Недавняя успешная космическая миссия ``Стардаст''
(Stardust) позволила впервые получить и привезти на Землю образцы межзвездной и кометной пыли, оставшейся со времен
образования Солнечной системы, анализ которых показал
присутствие силикатных веществ, кристаллизовавшихся при высокой ($\gtrsim800$~K) температуре, которые внедрены в
низкотемпературные конденсаты (\cite{Brownlee_ea04}). Присутствие кристаллических силикатов было
также обнаружено в протопланетных дисках вокруг других звезд (\cite{vanBoekel_ea04,Juhasz_ea10a}). Более того, в  составе
кометной пыли были обнаружены сложные органические соединения, включая простейшую аминокислоту глицин (\cite{Elsila_ea09}).
Изотопный анализ различных минералов в старейших метеоритах показал, что большинство элементов (за исключением
изотопов кислорода) обладают похожим изотопным составом, что предполагает эффективное перемешивание вещества во внутренней части
Солнечной туманности в течении первых нескольких миллионов лет ее существования (\cite{Boss2004}).

Все эти интересные факты отчасти объясняются современными моделями химической эволюции протопланетных дисков
(\cite{ah1999,vZea03,Woods_Willacy08,Visser_ea09,Semenov_Wiebe11a}).
Наиболее важный теоретический результат, частично подтвержденный наблюдениями, предсказывает ``слоистое'' распределение
молекул по диску. При этом в холодных внешних экваториальных частях дисков молекулы прилипают и постепенно вымерзают
на пыли, образуя протяженные ледяные мантии сложного состава, в то время как в ионизованной атмосфере диска молекулы
``разбиваются'' ультрафиолетовым излучением молодой и активной звезды. Также современные химические модели дисков позволяют
качественно и количественно объяснить наблюдаемые величины лучевых концентраций таких веществ, как
CO, HCO$^+$, N$_2$H$^+$, CN, HCN, HNC, CS.

\section{Молекулы как инструмент изучения протопланетных дисков}

\begin{table}
\caption{}
\label{obs_methods}
\vspace*{1.5ex}
\begin{center}
\begin{tabular}{llcccc}
\hline
{\bf Молекула}            & {\bf Параметр}     & {\bf Центральный}  & {\bf Молекулярный}    & {\bf Атмосфера}  & {\bf
Зона} \\
                          &                     & {\bf слой}  & {\bf слой}  &                         & {\bf планет} \\
\hline
$^{12}$CO, $^{13}$CO      & Температура        & мм$^{*}$   &  мм     & мм         &  ИК \\
H$_2$                     & Температура         & 0          &  0      & 0          &  ИК \\
NH$_3$                    & Температура        & см         &  см     & 0          &  0 \\
CS, H$_2$CO               & Плотность            & 0          &  мм     & 0          &  ИК  \\
CCH, HCN, CN              & Ионизирующее     & 0          &  мм     & 0          &  ИК  \\
                          & излучение        &&&&\\
HCO$^+$                   & Степень          & 0          &  мм     & 0          & 0   \\
N$_2$H$^+$                & ионизации        & мм         &  0      & 0          & 0   \\
C$^+$                     &                  & 0          &  0      & ИК         & ИК   \\
сложная                   & Поверхностные      & ИК$^{**}$  &  ИК-см  & 0          & ИК,мм \\
органика                  & процессы          &            &          &            &     \\
DCO$^+$, DCN,             & Изотопное          & мм         &   мм     &  0         & 0 \\
H$_2$D$^+$                & фракционирование      & & & & \\
\hline
\end{tabular}
\end{center}
\begin{flushleft}
$^{*}$ -- ``мм/см'' и ``ИК'' обозначают наблюдения в радио и инфракрасном диапазонах, соответственно.\\
$^{**}$ -- Сложная органика, вымороженная или образованная в ледяных мантиях пылинок, может быть обнаружена в виде слабых
линий поглощения на фоне сильного ИК-излучения центральной звезды. В газовой фазе ее легче всего наблюдать в
радио-диапазоне, на (суб-)миллиметровых длинах волн.\\
\end{flushleft}
\end{table}

К сегодняшнему моменту было найдено около $\sim 160$ химических соединений в космосе\footnote{\it{http://astrochemistry.net/}}.
Из них, в дисках, ввиду относительной малой массы излучающего газа, были обнаружены только лишь
CO (и его изотопологи), а также HCO$^+$, DCO$^+$, CN, HCN, DCN, CCH, H$_2$CO, и
CS (\cite{DGG97,Aikawa_ea03,Thi_ea04,Pietu_ea07,Henning_ea10}).
К тому же, соответствующие молекулярные спектры имеют, обычно, низкое соотношение сигнал/шум и пространственное разрешение.
Соответственно, радиальное распределение молекулярных обилий по дискам также определяется с большой
погрешностью (\cite{Pietu_ea05,Dutrey_ea07,Panic_ea09}). Путем тщательных и трудоемких наблюдений на таких первоклассных
инструментах, как интерферометр Плато де Бюр (Plateau de Bure interferometer) во Франции и Субмиллиметрового Интерферометра
(Submillimeter Array) в США, были изучены несколько наиболее крупных, ярких и близких к нам протопланетных дисков вокруг звезд
ДМ~Тельца (DM~Tau), Лкца~15 (LkCa~15), АБ~Возничего (AB~Aur) и ТВ~Гидры (TW~Hya). В 2013 году на полную мощность должен начать
работать Большой Миллиметровый Интерферометр в высокогорной пустыне Атакама, Чили (Atacama Large Millimeter Array; ALMA).
С вводом в строй этого инструмента, гораздо более чувствительного, обладающим разрешающей способностью вплоть до
0.005 угловых секунд, и способным наблюдать молекулы в широком диапазоне частот ($96-850$~ГГц) с высоким частотным разрешением,
изучение химического состава и физических условий в протопланетных дисках выйдет на принципиально другой уровень, позволяя
детально изучать большое количество объектов, включая маленькие и далекие системы, в линиях таких соединений о которых мы можем
сегодня только мечтать.

Методика изучения протопланетных дисков посредством наблюдений и моделирования эмиссионных линий молекул, обычно, начинается
с изучения физических условий в ярких линиях монооксида углерода, CO. Эти линии легко возбуждаются уже при таких низких
плотностях газа как $\sim 10^3-10^4$~см$^{-3}$, а потому в плотных протопланетных дисках заселенность уровней CO соответствует
ЛТР. Линии главного изотополога $^{12}$C$^{16}$O -- оптически толстые и наиболее яркие, поэтому измеряя их интенсивности можно
оценить кинетическую температуру газа в верхних областях дисков, а также поле скоростей, ориентацию, и геометрию (\cite{DGG97}).
Эмиссионные линии менее обильных изотопов, $^{13}$C$^{16}$O и $^{12}$C$^{18}$O, обычно, оптически тонкие или частично оптически
толстые. Их интенсивности чувствительны одновременно к температуре газа и поверхностной плотности этих молекул. Изучая кинематику
газа в этих линий, было установлено, что в дисках есть турбулентность, с
типичными дозвуковыми скоростями движений $\sim 0.05-0.2$~км\,с$^{-1}$ (\cite{DGH07,Hughes_ea11a}).

Было найдено, что размер дисков зависит от того, каким способом его оценивают: самые маленькие величины получаются
из наблюдений теплового излучения пыли, в то время как в линиях $^{12}$C$^{18}$O, $^{13}$C$^{16}$O, и
$^{12}$C$^{16}$O, диски кажутся все более и более большими, вплоть до радиусов в $300-1000$~а.е. Достоверно оценить радиус дисков
из наблюдений теплового излучения пыли пока не представляется возможным из-за того, что не хватает чувствительности
радиотелескопов. Для изотопологов CO разница в оценках размеров обусловлена частично нехваткой чувствительности телескопов, и
частично - селективной (изотопной) фотодиссоциацией, при которой основной изотополог, $^{12}$C$^{16}$O, разрушается медленнее,
чем менее распространенные изотопологи $^{12}$C$^{18}$O и $^{13}$C$^{16}$O. Также было получено, что большинство дисков обладают
градиентом кинетической температуры газа в вертикальном направлении, как и предсказывается современными моделями, с температурами
около 10~K в центральной области и 50~K -- в молекулярном слое 
(\cite{DDG03,Pietu_ea07,Isella_ea10a}). Есть и исключения, а именно сильно проэволюционировавшие диски с более однородным
распределением температуры из-за их особенной физической структуры, в который одновременно присутствуют маленькие пылинки во
внешней части, и большие пылинки (а может, уже и планетозимали) во внутренней части, например, ГМ~Возничего (GM~Aur) и
Лкца~15 (LkCa~15) (\cite{Dutrey_ea08,Hughes_ea09}). В одном массивном и протяженном диске у ДМ~Тельца был обнаружено присутствие
очень холодных ($\sim 10-15$~K) газов, таких как CO, CCH, HCO$^+$, CN and HCN. Данный факт пока не может быть достоверно объяснен
в рамках современных астрохимических моделей.

Для молекул с большим дипольным моментом, чем у CO (0.112~Дебая) требуется более высокая плотность газа для возбуждения
вращательных переходов. После линий монооксида углерода, HCO$^+$ (с дипольным моментов 3.92~Дебая)--
вторая наиболее легко наблюдаемая химическая специя в дисках. Вращательные
переходы этого иона термализуются уже при плотностях газа около 10$^5$~см$^{-3}$, и чувствительны к плотности газа. Это один из
самых распространенных ионов в протопланетных дисках (другие ионы, не наблюдаемые в радио-диапазоне, это C$^+$ и H$_3^+$).
Иногда в дисках удается отнаблюдать линии N$_2$H$^+$ (дипольный момент 3.37~Дебая). Используя наблюдения этих двух ионов было
установлено, что степень ионизации молекулярного слоя диска составляет $\sim 10^{-10}-10^{-9}$, как и предсказывается химическими
моделями. Этой величины
достаточно для эффективного взаимодействия газа и магнитного поля и, т.о., для образования турбулентности в дисках
(\cite{Dutrey_ea07}).

Линии остальных молекул и ионов еще менее яркие, и требуют большого количества наблюдательного времени даже на самых
чувствительных интерферометрах. Например, было установлено, что отношение интенсивностей линий C$_2$H и CN к HCN
зависит от интенсивности и формы УФ спектра звезды (\cite{BCDH03}). Т.е., чем выше интенсивность УФ излучения звезды, тем
более обильными становятся радикалы  C$_2$H и CN. 
DCO$^+$ и DCN имеют относительно высокие концентрации по сравнению с основными изотопологами HCO$^+$ и HCN,
$\sim 1-10\%$, хотя космическое отношение элементов D/H составляет только около $10^{-5}$ (\cite{Qi_ea08}). Пока непонятно, что
обуславливает такую высокую степень фракционирования дейтерия в
этих молекулах: с одной стороны, это может быть  ``наследие'' химической эволюции в холодных молекулярных облаках,
или же такие процессы могут эффективно протекать в холодных частях дисков.

Один из наиболее важных наблюдательных результатов касающихся химического состава протопланетных дисков состоит в том, что
измеренные обилия молекул оказываются систематически в 5--100 раз ниже, чем в молекулярных облаках в области
мало-массивного звездообразования в Тельце (\cite{DGH07}). Это объясняется более сильным вымораживанием газофазных молекул
в холодных и плотных центральных областях дисков, а также их быстрой фотодиссоциацией в сильно ионизированной атмосфере.
Присутствие большого количества льдов в холодных частях дисков было обнаружено с помощью Инфракрасной Космической Обсерватории
(Infrared Space Observatory; ISO) и телескопа ``Спитцер'' (\cite{vanDishoeck_ARA2004,Bouwman_ea08}). В основном, они
состоят из воды с примесью ($\sim 1-30$\%) более сложных соединений, например, CO, CO$_2$,
NH$_3$, CH$_4$, H$_2$CO, and HCOOH (\cite{Zasowski_ea08}). С помощью ``Спитцера'' были найдены
колебательно-вращательные и колебательные линии таких молекул как CO, CO$_2$, C$_2$H$_2$, HCN, H$_2$O и OH, которые для
возбуждения требуют температур выше $\sim 300-1\,000$~K. Было обнаружено, что отношение интенсивностей линий HCN к C$_2$H$_2$
чувствительно к интенсивности ионизирующего излучения звезды. С помощью телескопа ``Гершель''
начались подробные исследования химического состава атмосферы дисков, в частности, путем наблюдений инфракрасных переходов
CII (158~мкм) и ОI (63~мкм), а также линий легких ионов (CH$^+$, OH$^+$, H$_2$O$^+$, HS$^+$), см \cite{Mathews_ea10,Thi_ea11a}.

В таблице~\ref{obs_methods} приведены наблюдаемые молекулярные линии, которые используются для исследования
физических условий в разных областей протопланетных дисков.

\section{Общая схема химической структуры протопланетного диска}

\begin{figure}
\includegraphics[angle=0,width=9cm]{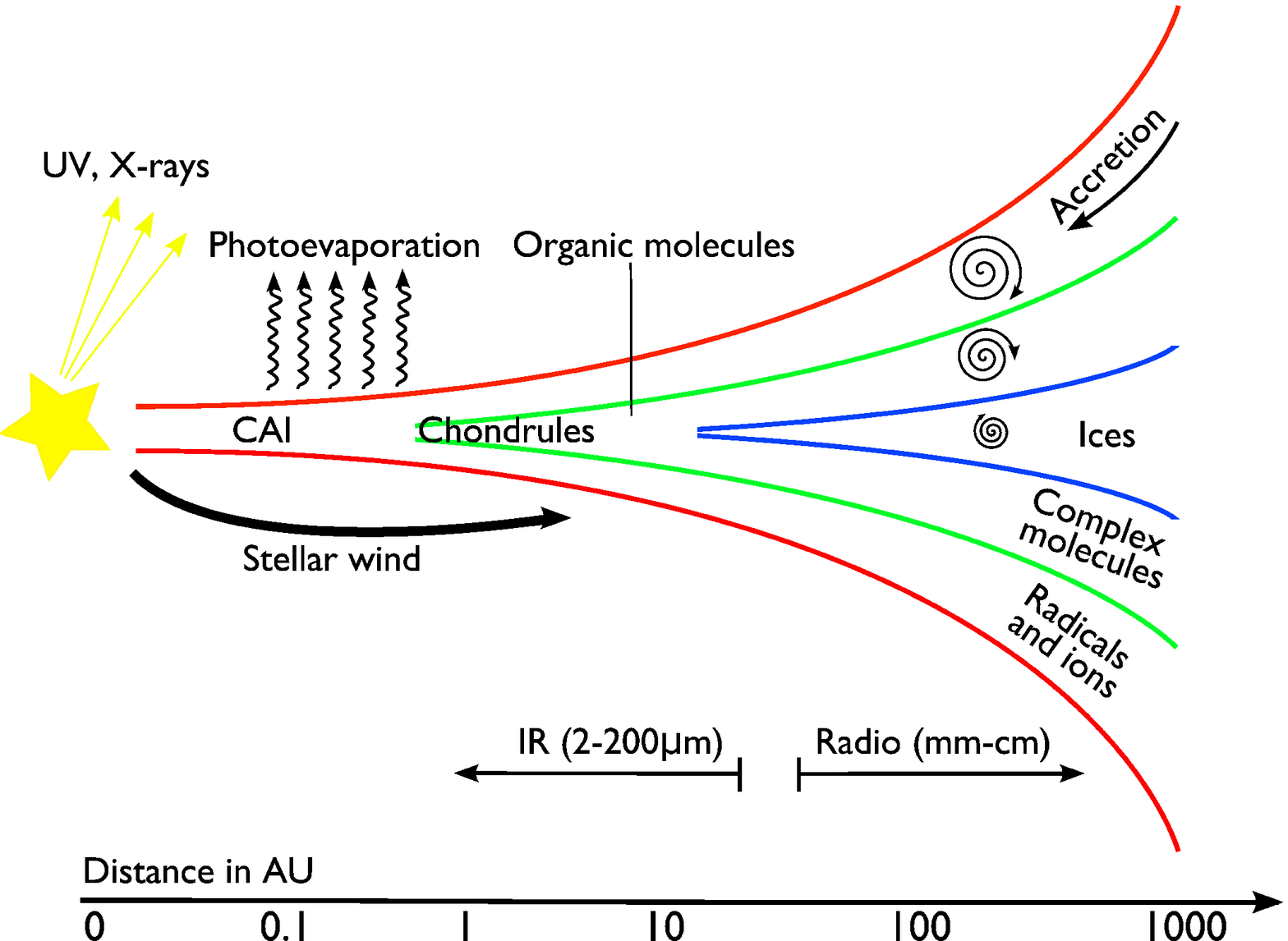}
\caption{Схема предполагаемой физической и химической структуры протопланетного диска вокруг молодой мало-массивной звезды
типа Солнца (расстояния показаны не в масштабе). Рисунок позаимствован из работы \cite{Semenov_10}.
\label{scheme}}
\end{figure}

Общая схема структуры протопланетного диска у мало-массивной звезды показана на рисунке~\ref{scheme}.

Перенос и сохранение углового момента в протопланетном диске за счет турбулентной вязкости газа обуславливает его
динамическую ( ``вязкую'') эволюцию. При этом значительная доля вещества ($\sim 80\%$) переносится к звезде, а остальное вещество
уносит угловой момент наружу. Этот процесс приводит к тому, что радиальное распределение поверхностной плотности по диску может
быть с хорошей точностью описано степенным законом, с показателем степени около -3/2. Для модели Солнечной туманности
\cite{mmsn} поверхностная плотность на 1~a.e. составляет около $1\,700$~г\,см$^{-2}$. В вертикальном направлении, плотность диска
падает экспоненциально, в то время как температуры пыли и газа растут, причем, начиная с некоторой высоты газ становиться горячее
чем пыль (из-за менее эффективного охлаждения и нагрева интенсивным излучением центральной звезды). Также, температура возрастает
по направлению к звезде, причем, на расстояниях $\lesssim0.1$~a.e., где температура становится выше $1\,500-2\,000$~K, начинается
испарение пыли. Испарение льдов начинается гораздо раньше, на расстоянии $\sim 3$~а.е. ($T \sim 100-150$~K) испаряется водяной
лед, а на $\sim 20-50$~а.е. ($T \sim 20$~K) испаряется СО лед.

Проникновение в диск ионизирующего УФ и рентгеновского
излучения звезды определяется поглощающими и рассеивающими свойствами пыли, а также распределением плотности газа, и энергией
фотона/частицы. Так, например, УФ излучение почти полностью блокируется лучевой концентрацией вещества $\lesssim0.01$~г~см$^{-2}$
(атмосфера и верхняя часть молекулярного слоя), в то время как для жестких рентгеновских фотонов ($1-5$~кэВ) эта величина
составляет $0.1-1$~г~см$^{-2}$ (атмосфера и молекулярный слой), а для релятивистских частиц космических лучей --
$100$~г~см$^{-2}$ (практически весь диск).

Соответственно, с химической точки зрения, весь диск может быть поделен на 4 основные области (характеризуемые, в основном,
различной температурой и интенсивностью ионизирующего излучения). Внутренняя область соответствует
области образования планетной системы (с радиусами $\lesssim 20-50$~a.e), которая доступна для наблюдений с помощью
высококлассных
инфракрасных телескопов (и, в будущем, с ALMA). Внешняя область диска ($\gtrsim20-50$~a.e) доступна для наблюдений в
радио-диапазоне с современными интерферометрами и, в свою очередь, состоит из 3 вертикальных слоев:
экваториального (центрального), теплого молекулярного слоя и атмосферы.

Плотная ($>10^8$~см$^{-3}$) экваториальная часть полностью непрозрачна для внешнего ионизирующего излучения (ИК, рентген) из-за
поглощения
пылью, а потому остается холодной ($10-20$~K) и практически химически нейтральной. Химические процессы в этой части диска
сначала протекают в газовой фазе, за счет быстрых ион-молекулярных реакций, но, вследствие постепенного вымораживания
молекул на пыль основную роль начинают играть медленные поверхностные реакции (в основном, с участием радикалов и атомов
водорода). Молекулы, образованные на или прилипшие к пылинкам, едва ли могут десорбировать обратно в газ. Характерное время
химических процессов в холодной центральной области диска определяется медленными поверхностными реакциями и составляет
около $>10^5-10^6$~лет (для радиусов $\sim 100-300$~а.е.). Ближе к звезде, центральная область диска (``внутренняя область'')
становится горячее ($>100$~K), в том числе и из-за аккреционного нагрева, что сводит роль химии льдов на нет, и характерное
химическое время становится коротким ($\sim$100~лет). Поэтому в той части диска, где образуются планеты, химические процессы
находятся в равновесии, что позволяет серьезно упростить их моделирование путем использования термодинамических моделей.

Сразу над центральной областью находится более теплый ($T\sim30-70$~K), менее плотный ($\sim 10^6-10^7$~см$^{-3}$),
частично облучаемый ионизирующим излучением, молекулярный слой. Химическая эволюция в этой части диска определяется, в основном,
эффективными взаимодействиями между процессами, протекающими в газе и на поверхности пыли, и регулируется УФ и рентгеновским
излучением центральной звезды (\cite{Bergin_ea07}). Образование сильного УФ-поля нетепловой природы в звездах типа Т~Тельца
связано с их повышенной хромосферной активностью и аккрецией вещества диска на звезду
(\cite{Bouvier_ea07}). Подобным же образом генерируется низко-энергетическое рентгеновское излучение, в то время как более
жесткий рентген образуется в горячих аккреционных джетах и за счет вспышечной активности молодой звезды
(наподобие солнечной активности, только в тысячи раз более сильной), см. \cite{Guedel_Naze09}.
Рентгеновские и УФ-фотоны ионизируют газ, приводят к диссоциации молекул в газе и выбивают их с поверхности пылинок.
Повышенное количество сложных ионов и радикалов, а также постоянный обмен веществом между газом и ледяными мантиями
обуславливают богатый химический состав теплого молекулярного слоя. В нем относительные концентрации многих молекул достигают
своего пика, и большинство эмиссионных линий образуется именно здесь. Из-за большого числа разнообразных химических процессов,
активных в молекулярном слое, время достижения химического равновесия сопоставимо с, или даже превышает жизни протопланетного
диска, $>1$ миллиона лет.

Наконец, над молекулярным слоем находится горячая ($T\sim100-10\,000$~K), разреженная ($\lesssim 10^5$~см$^{-3}$), и сильно
ионизованная атмосфера диска. В ней практически нет никаких молекул и сложных ионов, за исключением молекулярного водорода
и таких фото-стабильных радикалов, как CCH и CN. Химические процессы в атмосфере протекают исключительно в газе, и состоят из
фото-реакций и реакций диссоциативной рекомбинации. Характерное время химической эволюции атмосферы диска составляет всего
около $\sim$100~лет (для $\sim 100-300$~а.е.).

В следующих разделах обзора более подробно рассказывается о разных типах химических процессов, активных в 4 основных областях
протопланетных дисков.

\section{Газофазная химия}
Современные астрохимические модели содержат данные о тысячах реакций, которые могут быть активны в космических условиях
(\cite{OSU03,Woodall_ea07,Wakelam_09b}). К сожалению, только лишь $10-20\%$ из них хорошо изучены в лаборатории или
путем сложных квантово-механических расчетов. Поэтому не стоит забывать о том, что результаты любого астрохимического
моделирования обладают внутренне-обусловленными неточностями с ошибками в расчетных концентрациях в несколько раз
(\cite{Vasyunin_ea04,Vasyunin_ea08}).
Все эти тысячи реакций могут быть поделены на 4 основные группы, которые соответствуют 4 химическим регионам диска
(см. таблицу~\ref{bond}, адаптированную из \cite{vanDishoeck88}).

\begin{landscape}
\begin{table}
\caption{Типы химических процессов, активных в протопланетных дисках}
\label{bond}
\vspace*{1.5ex}
\centering
\scriptsize
\begin{tabular}{llcccc}
\hline
{\bf Процесс} & {\bf Формула} & {\bf Центральный} & {\bf Молекулярный}   & {\bf Атмосфера}   & {\bf Зона} \\
              &               & {\bf слой}          & {\bf слой}            &                   & {\bf планет} \\
\hline
{\bf Образование  связей} & & & & & \\
Радиативная ассоциация  & A + B $\rightarrow$ AB + h$\nu$          & X         &  X    & X   &  X  \\
Поверхностная реакция      & A + B$\|$gr    $\rightarrow$ AB + gr    &  X         & X    & 0   &  0 \\
Трех-частичная реакция            & A + B + M $\rightarrow$ AB + M          &  0         & 0    & 0   & X \\
\hline
{\bf Разрушение  связей} & & & & & \\
Фотодиссоциация      & AB + h$\nu$ $\rightarrow$ A + B          &  0         & X    & X   & X \\
Диссоциация космическими & AB + CRP $\rightarrow$ A + B                & X           & X    & 0    & 0 \\
лучами & & & & & \\
Диссоциация рентгеном & ---                                      & 0           & X    & X    & X \\
Диссоциативная           & AB$^+$ + e$^-$ $\rightarrow$ A + B       & X          & X    & X   & X  \\
рекомбинация          & & & & & \\
\hline
{\bf Обмен связями} & & & & & \\
Нейтраль-нейтральная &      A + BC $\rightarrow$ AB + C                & X         & X     & 0   & X \\
реакция & & & & & \\
Ион-молекулярная реакция         & A$^+$ + BC $\rightarrow$ AB$^+$ + C      &  X         & X     & X   & X \\
Перераспределение заряда        & A$^+$ + BC $\rightarrow$ A + BC$^+$     &  X         & X     & X   & X \\
\hline
{\bf Сохранение связей} & & & & & \\
Фотоионизация         & AB + h$\nu$ $\rightarrow$ AB$^+$ + e$^-$ & 0       & X     & X   & X \\
Ионизация космическими & AB + CRP $\rightarrow$ AB$^+$ + e$^-$           & X       & X     & 0   & 0 \\
лучами & & & & & \\
Ионизация рентгеном  & ---                                       & 0       & X     & X   & X \\
\hline
\end{tabular}
\end{table}
\end{landscape}

Почти во всем диске большинство активных химических реакций -- это кинетические процессы первого (например, молекула плюс фотон)
или второго (например, ион плюс молекула) порядков. В самой внутренней части диска, при плотностях, превышающих $\sim
10^{10}$~см$^{-3}$, становятся активными также трех-частичные процессы (например, молекула плюс молекула плюс молекула).
Основными процессами, благодаря которым начинается образование молекулярных связей, являются медленные реакции радиативной
ассоциации и поверхностные реакции. С некоторой вероятностью столкновения в газе могут привести к образованию так называемого
столкновительного комплекса, находящегося в возбужденном состоянии, который либо стабилизируется путем переизлучения избытка
энергии фотоном(-ами), либо снова распадается (\cite{HerbstKlemperer73}). Например, образование углеродных
цепочек начинается с радиативной ассоциации между C$^+$ и H$_2$ в CH$_2^+$ (\cite{Herbst85}).

Далее, вновь образованные молекулы и ионы начинают реагировать друг с другом посредством быстрых ион-молекулярных реакций.
Эти реакции экзотермичны, т.е. протекают с выделением тепла и не обладают энергетическими барьерами, и обладают большими
коэффициентами скоростей, $\sim 10^{-9}$~см$^{-3}$\,с$^{-1}$, которые зачастую увеличиваются при уменьшении температуры
за счет Кулоновского притяжения между ионом и центром заряда молекулы (\cite{DalgarnoBlake76}). Ион-молекулярные реакции
приводят к перераспределению молекулярных связей между реактантами. Этот тип химических процессов составляет наибольшую долю
реакций в астрохимических моделях.

С другой стороны, молекулярные ионы могут сталкиваться с электронами или заряженными пылинками, диссоциативно рекомбинировать,
и распадаться на несколько фрагментов. Эти процессы тоже проходят с выделением энергии, которая превращается в кинетическую
энергию продуктов диссоциации. Типичные константы скоростей реакций диссоциативной рекомбинации составляют
около $10^{-7}$~см$^{-3}$\,s$^{-1}$ при $10$~K (\cite{Woodall_ea07}). Практически для всех наблюдаемых молекул в дисках
реакции диссоциативной рекомбинации являются важным каналом их образования (например, для воды и углеводородных цепочек).
Зачастую, на поздних этапах эволюции диска, $\gtrsim 10^5$~лет, скорость разрушения многоатомных ионов за счет диссоциативной
рекомбинации уравновешивается реакцией протонирования, т.е. добавлением протона за счет ион-молекулярной реакции с H$_3^+$.
Например, процесс CO + H$_3^+$ $\rightarrow$ HCO$^+$ + H$_2$ уравновешивается реакцией HCO$^+$ + e$^-$
$\rightarrow$ CO + H. К сожалению, не так просто точно теоретически предсказать или измерить в лаборатории то соотношение
обилий разных фрагментов, на которое распадается сложный молекулярный ион (\cite{SmithSpanel94}).

Относительно недавно было показано, что некоторое количество нейтраль-нейтральных реакций, включающих радикалы, могут быть
активными в космических условиях (\cite{vanDishoeckxx}). Константы скоростей этих процессов
составляют $\sim10^{-11}$--$10^{-10}$~см$^{-3}$\,s$^{-1}$, что всего лишь на несколько порядков меньше, чем в случае быстрых
ион-молекулярных реакций (\cite{Clary85}). Одна из наиболее важных реакций такого типа в протопланетных дисках приводит к
образованию HCO$^+$ из O и CH.

\section{Фотохимические процессы}
Молодые звезды типа Т~Тельца ($T_{\rm eff} \simeq 4\,000$~K) обладают мощным УФ полем, спектр которого отличается от спектра
межзвездного УФ излучения, в частности, наличием сильной эмиссионной линии Лайман-альфа ($121.6$~нм; \cite{BCDH03}). Интегральная
интенсивность УФ излучения на расстоянии 100~а.е. от такой звезды может более чем в сотни раз превышать интенсивность
межзвездного УФ поля (\cite{Habing68}). Разрушение молекул ультрафиолетовыми фотонами может происходить несколькими путями, в
частности, либо за счет поглощения фотонов определенных энергий, либо за счет поглощения в континууме, либо сразу за счет
обоих механизмов. При этом важную роль играет распределение поглощающей УФ пыли (суб-)микронного размера по диску, которое может
значительно изменить эффективность проникновения ионизирующего излучения во внутренние части (\cite{Vasyunin_ea11}).
Например, такие важные для изучения дисков как CO, H$_2$, и CN, разрушаются путем поглощения УФ-фотонов
определенных энергий, $\lambda \lesssim 1\,100$~{\AA}, а потому не чувствительны к наличию (или отсутствию) мощной эмиссионной
линии Лайман-альфа в УФ-спектре. Другие специи, например, HCN и CH$_4$, разрушаются УФ-фотонами меньших энергий, поэтому
эффективность их фотодиссоциации зависит от интенсивности линии Лайман-альфа (\cite{vanDishoeck88,vDea_06}). Именно
этим фактом объясняется наблюдаемое повышенное отношение концентрации CN к HCN в дисках (\cite{DGG97,BCDH03}).

Так как диссоциация чрезвычайно распространенных молекул водорода и монооксида углерода происходит путем взаимодействия с
УФ-фотонами определенных (высоких) энергий, ее эффективность зависит не только от количества поглощающих пылинок в среде, но и от
количества самих H$_2$ и CO на пути распространения УФ излучения. Каждый процесс разрушения этих молекул приводит к ``выеданию''
энергии из УФ спектра на определенных частотах, что, хоть и мало сказывается на полной интенсивности ультрафиолетового
излучения, уменьшает вероятность диссоциации в последующих случаях (механизм самозащиты от фотодиссоциации, self-shielding). Для
H$_2$ и CO критическим лучевыми концентрациями газа для практически полной остановки фотодиссоциации являются
$\sim 10^{14}$~см$^{-2}$ и $\sim 10^{21}$~см$^{-2}$, соответственно (\cite{DB96,Visser_ea09b}).

Однако, это условие выполняется только для главных изотопологов H$_2$ и $^{12}$C$^{16}$O. Менее распространенные изотопологи
HD, D$_2$, $^{13}$C$^{16}$O, $^{12}$C$^{18}$O, $^{13}$C$^{18}$O, $^{12}$C$^{17}$O диссоциируют путем поглощения УФ фотонов
немного других энергий, а потому механизм самозащиты от фотодиссоциации в дисках для них не работает (или эффективен лишь
частично). Таким образом, получается, что в молекулярном слое и атмосфере диска соотношение обилий атомов $^{12}$C и
$^{13}$C, а также изотопов кислорода может отличаться от изначального космического, что приводит к изотопно-селективной химии и
измененным концентрациям изотопных соединений (\cite{vanDishoeck88,Lea96,Visser_ea09b}).
Именно такой изотопно-селективной УФ-диссоциацией (вкупе с динамической эволюцией) объясняется наличие аномальных соотношений
изотопов кислорода в различных минералах, образованных на стадии образования Солнечной системы, и входящим в состав метеоритного
вещества (\cite{Lyons_Young05}).

В следующем разделе вкратце описываются взаимодействие молекул с пылью и поверхностная химия. Более подробное описание приведено
в главе А.~И.~Васюнина в этом сборнике.

\section{Взаимодействие газа с пылевыми частицами и поверхностные реакции}

\begin{table}
\caption{Энергии десорбции для астрохимически важных молекул}
\label{des_energies}
\vspace*{1.5ex}
\begin{center}
\begin{tabular}{ll}
\hline
{\bf Молекула}            & {\bf Энергия десорбции, К}  \\
\hline
 C$_2$H            &  2140$^{*}$ \\
 C$_2$H$_2$           &  2590 \\
 C$_2$S            &  2700 \\
 C$_3$H$_2$           &  3390 \\
 C$_6$H$_6$           &  7590 \\
 CH$_2$CO         &  2200 \\
 CH$_3$CHO         &  2870 \\
 CH$_3$OH          &  5530 \\
 CH$_4$            &  1300 \\
 CN             &  1600 \\
 CO             &  1150 \\
 CO$_2$            & 2580 \\
 CS             &  1900 \\
 H              &  624 \\
 H$_2$             &  552 \\
 H$_2$S             & 2740 \\
 H$_2$CO           &  2050 \\
 H$_2$O            &  5700 \\
 HCN            &  2050 \\
 HCOOH          &  5570 \\
 HNC            &  2050 \\
 HNCO           &  2850 \\
 HNO            &  2050 \\
 N              &  800  \\
 N$_2$             &  1000 \\
 NH$_3$            &  5530 \\
 NO             &  1600 \\
 O              &  800 \\
 O$_2$             &  1000 \\
 OH             &  2850  \\
 S              &  1100  \\
 SO             &  2600  \\
 SO$_2$            &  3400 \\
\hline
\end{tabular}
\end{center}
\begin{flushleft}
$^{*}$ -- примерная температура, при которой начинается испарение молекулы из ледяной мантии, может быть получена путем деления
энергии десорбции (в К) на 50. \\
\end{flushleft}
\end{table}

В холодных ($\lesssim10-100$~K) областях протопланетных дисков многие молекулы успевают почти полностью осесть на пылевые
частицы за несколько миллионов лет эволюции. При таких низких температурах молекулы прилипают (адсорбируют) при столкновениях к
пылинкам почти со 100\% вероятностью, так как их кинетические энергии ($\lesssim 100$~K) намного меньше энергий испарения
(десорбции; $\sim 600-6\,000$~K), см. таблицу~\ref{des_energies} (\cite{dHendecourtea85}).
Точно оценить вероятность сложно, так как она зависит от свойств поверхности (пористость, материал, поверхностный потенциал).
Прилипание к
поверхности происходит не хаотично, а в определенные участки с пониженным электростатическим потенциалом (участки адсорбции).
На пылинке с радиусом 0.1 мкм состоящей преимущественно из аморфных силикатов, размещается около миллиона таких участков
(со средним размером в несколько Ангстрем).

Прилипание молекул может происходить двумя различными способами, за счет физической адсорбции (физисорбции) и химической
адсорбции (хемосорбции). В первом случае молекула удерживается на поверхности за счет сил ван дер Ваальса, а во втором -- за счет
образования химической связи с молекулой поверхности. Энергия прилипания для физической адсорбции составляет всего лишь доли эВ
(что соответствует $600-6\,000$~K, см. таблицу~\ref{des_energies}), в то время как энергия химической адсорбции гораздо выше,
$\gtrsim1$~эВ ($\gtrsim10-30\,000$~K).
Также как и молекулы, к пылинкам могут прилипать электроны, что делает возможным
реакции диссоциативной рекомбинации ионов на пыли. В самой темной, плотной и почти нейтральной части центральной области диска
отрицательно заряженные пылевые частицы становятся наиболее распространенными носителями заряда (\cite{Red2}).

Помимо адсорбции, молекулы могут, при благоприятных условиях, десорбировать с поверхности. Конечно, для тех молекул, которые
хемосорбировались на поверхность пылинки вероятность десорбции обратно в газовую фазу невелика. Считается, что в протопланетных
дисках реализуются три наиболее эффективных механизмов десорбции (отлипания) льдов. Это тепловая десорбция, десорбция за счет
нагрева пылинки релятивистской частицей космических лучей, и десорбция ультрафиолетовыми фотонами.

Тепловая десорбция начинается, когда поверхностные молекулы нагреваются так, что их кинетическая энергия начинает превышать
энергию прилипания (или энергию десорбции). Легкие простые молекул, такие как CO и N$_2$, начинают десорбировать при нагреве
пылинки до 20~K (\cite{Bisschop_ea06}), в то время как для воды, диоксида углерода и тяжелых соединений (углеродных цепочек,
цианополиинов, органики) требуются температуры превышающие $\sim 50-150$~K. Хемосорбировавшие молекулы начинают испарятся только
при достижении температур $\gtrsim300-1\,000$~K. Очевидно, что тепловая десорбция очень важна для химической эволюции внутренней
($\lesssim 1-10$~a.e.) горячей области протопланетных дисков.

Во внешней холодной центральной области дисков работает другой механизм десорбции -- десорбция за счет взаимодействия космических
лучей с веществом. В состав космических лучей входит небольшая ($<1\%$) часть ядер железа релятивистских энергий ($>1$~ГеВ).
Если такая высокоэнергичная частица сталкивается с пылинкой, то ее энергии хватает, чтобы  нагреть последнюю до высокой
температуры, $\gtrsim 50$~K, позволяя импульсно испарить часть или всю ледяную мантию (\cite{WatsonSalpeter72,Leger_ea85}).
Также частицы космических лучей (или ``выбитые'' ими быстрые электроны), сталкиваясь с молекулами водорода, вызывают их
возбуждение, заставляя излучать УФ фотоны, которые могут, в свою очередь, испарять поверхностные молекулы фотодесорбцией
(\cite{PrasadTarafdar83}).

Фотодесорбция поверхностных специй важна для химических процессов в молекулярном слое дисков. При этом каждый УФ фотон, попавший
на поверхность ледяной мантии пылинки, может с какой-то вероятностью ``отколоть'' поверхностную специю. За последние несколько
десятилетий лабораторные исследования позволили измерить эти вероятности для дюжины наиболее важных простых молекул. Эта
вероятность (на каждый УФ фотон) для CO, H$_2$O, CH$_4$ и NH$_3$ составляет около $10^{-4}-10^{-2}$, и зависит от формы УФ спектра
(\cite{Oeberg_ea09a,Oeberg_ea09b,Fayolle_ea11a}).

Наконец, молекулы на поверхности при определенных условиях могут вступать в химические реакции друг с другом. При этом
поверхность пылинки выступает в роли своеобразного катализатора реакций, поглощая выделяющееся тепло и позволяя
реагентам накапливаться. Самая важная молекула во Вселенной, H$_2$, образуется из атомарного водорода практически всецело на
поверхности пыли (\cite{HollenbachSalpeter71,WatsonSalpeter72}). Поверхностные реакции могут протекать несколькими способами.
Если реагирующие молекулы физисорбированы, то при определенном диапазоне температур они могут перескакивать (или туннелировать в
случае H и H$_2$) в соседние участки адсорбции, так как потенциальные барьеры между ними обычно меньше, чем энергия десорбции.
Таким образом, физисорбированные реагенты могут блуждать по поверхности и, в конце концов, прореагировать друг с другом, если
обе молекулы оказываются на одном и том же участке адсорбции. Этот механизм химических реакций  на поверхности называется
механизмом Ленгмюра-Хиншельвуда, и в дисках он работает в центральной темной области.
Существует еще один важный  механизм химических реакций на поверхности, механизм Элей-Редила,
когда один из радикалов хемосорбирован на поверхности, а другая газофазная молекула сталкивается с ним, образуя с некоторой
вероятностью новую специю. Этот механизм особенно важен для углеродистых поверхностей и играет важную роль в поверхностной химии
при высоких ($\gtrsim100$~K) температурах и/или интенсивной фотодиссоциации, т.е., в молекулярном слое диска.
Лабораторные исследования показали, что сложные молекулы, включая органику (например, метанол), могут эффективно образовываться в
космических условиях в основном за счет поверхностной химии (\cite{Geppert_ea05}).


\section{Заключение}
В данной обзорной лекции показано как образуются и разрушаются молекулы в процессе эволюции протопланетных
дисков -- аналогах нашей Солнечной системы, когда ей было всего несколько миллионов лет. Молекулы являются важными инструментами
изучения физических условий, кинематики, и химической структуры дисков. На протопланетной стадии эволюции , сильные
градиенты плотности, температуры, и интенсивности ионизирующего излучения приводят к образованию ``слоистой'' химической
структуры. В горячей, разреженной и сильно ионизованной атмосфере присутствуют только лишь атомы, атомарные ионы, и простые
фотостабильные радикалы, чья химическая эволюция определяется ограниченным набором газофазных реакций. В более глубоких слоях
диска рентгеновское и ультрафиолетовое межзвездное и звездное излучение начинает ослабляться за счет поглощения пылью и газом,
температура понижается, позволяя образовываться множеству сложных соединений посредством газопылевой химии. Наконец, в темной,
холодной и плотной экваториальной части диска большинство молекул успевает вымерзнуть за время жизни диска, образуя сложные
ледяные мантии на пылинках, на которых происходит образование сложных (органических) молекул за счет поверхностных реакций.
Вкратце рассказано, какие основные  факты о структуре дисков были получены в результате наблюдений на радио и инфракрасных
телескопах.

\vspace{5mm} \small
Работа выполнена при поддержке гранта Немецкого Научного Фонда в рамках приоритетной программы
SPP~1385: ``The first ten million years of the Solar system - a
planetary materials approach'' (SE 1962/1-1).
\normalsize

\input journal.tex
\bibliography{references}

%% file: journal.tex
\def\apj{Astrophys.~J}
\def\aatr{Astron.~Astroph.~Trans}
\def\aaps{Astron.~and Astrophys.~Suppl.~Ser}
\def\pasp{Publ.~Astron.~Soc.~Pac}
\def\gca{Geochim.~Cosmochim.~Acta}
\def\aap{Astron.~Astrophys}
\def\aspcs{ASP~Conf.~Ser}
\def\asrep{Astron.~Rep}
\def\nat{Nature}
\def\apjl{Astrophys.~J.,~Lett}
\def\apjs{Astrophys.~J.,~Suppl.~Ser}
\def\aj{Astron.~J}
\def\mnras{Mon.~Not.~R.~Astron.~Soc}
\def\araa{Ann.~Rev.~Astron.~Astrophys}
\def\jcp{J.~Chem.~Phys}
\def\apss{Astrophys.~Space.~Sci}
\def\prl{Phys.~Rev.~Lett}
\def\phrva{Phys.~Rev.~A}
\def\phlb{Phys.~Let.~B}
\def\pf{Phys.~Fluids}
\def\azh{Астрон.~журн}
\def\pazh{Письма~в~Астрон.~журн}
\def\jgr{J.~Geophys.~Res}
\def\cemda{Celest.~Mech.~Dyn.~Astr}
\def\jcoph{J.~Comp.~Phys}
\def\cophc{Comput.~Phys.~Commun}
\def\phpl{Physics~of~Plasmas}
\def\pasj{Publ.~Astron.~Soc.~Jpn}
\def\avest{Астрон.~вест}
\def\jrasc{J.~R.~Astron.~Soc.~Can}
\def\cemec{Celest.~Mech}
\def\pasau{Proc.~Astron.~Soc.~Aust}
\def\puasau{Publ.~Astron.~Soc.~Aust}
\def\jasa{J.~Acoust.~Soc.~Am}
\def\jfm{J.~Fluid~Mech}
\def\cajph{Can.~J.~Phys}
\def\mitag{Mitt.~Astron.~Ges}
\def\bain{Bull.~Astron.~Inst.~Neth}
\def\epsl{Earth~Planet.~Sci.~Lett}
\def\ibvs{Inf.~Bull.~Variable~Stars}
\def\arep{Astr.~Rep}
\def\phr{Phys.~Rep}
\def\astl{Astron.~Letters}
\def\sci{Science}
\def\jqsrt{J.~Quant.~Spectrosc.~Radiat.~Transfer}
\def\emp{Earth,~Moon~and~Planets}
\def\icar{Icarus}
\def\pss{Planet.~Space~Sci}
\def\qjras{Q.~J.~R.~Astron.~Soc}
\def\nimpa{Nucl.~Instrum.~Methods~Phys.~Res.,~Sect.~A}
\def\soph{Sol.~Phys}
\def\lnm{Lect.~Notes~in~Math}
\def\an{Astron.~Nach}
\def\aph{Astroparticle~Physics}
\def\adspr{Adv.~Space~Res}
\def\geoj{Geophys.~J}
\def\caosp{Contrib.~Astron.~Obs.~Skalnat{\'e}~Pleso}
\def\vestcpbu{Вестн.~С.-Петерб.~ун-та}
\def\izvvrad{Изв.~вузов.~Радиофизика}
\def\izvans{Изв.~AH~CCCP}
\def\vestvgu{Вестн.~ВолГУ}
\def\bamass{Bull.~Am.~Astron.~Soc}
\def\rmxaa{Rev.~Mex.~Astron.~Astrofis}
\def\aapr{Astron.~Astrophys.~Rev}
\def\acp{Atmosphere~Chem.~Phys}
\def\cosiss{Космич.~исслед}
\def\ssrv{Space~Sci.~Rev}
\def\jmph{J.~Math.~Phys}
\def\rvmps{Rev.~Mod.~Phys.~Suppl}
\def\rvmp{Rev.~Mod.~Phys}
\def\prd{Phys.~Rev.~D}
\def\nuphs{Nuc.~Phys.~B~Proc.~Suppl}
\def\nuphb{Nuc.~Phys.~B}
\def\skytel{Sky~Telesc}
\def\thmc{Тез.~международ.~конф.}
\def\mmc{Материалы~международ.~конф.}
\def\mvrc{Материалы~всерос.~конф.}
\def\cntc{Сб.~науч.~тр.~конф.}
\def\tmnpc{Тр.~Международ.~науч.-практ.~конф.}
\def\ctc{Сб.~тр.~конф.}
\def\mcnc31{Тр.~31-й~Международ.~студ.~науч.~конф., Екатеринбург, 28 янв.---1 февр. 2002~г}
\def\mcnc32{Тр.~32-й~Международ.~студ.~науч.~конф., Екатеринбург, 3---7 февр. 2003~г}
\def\mcnc33{Тр.~33-й~Международ.~студ.~науч.~конф., Екатеринбург, 2---6 февр. 2004~г}
\def\mcnc34{Тр.~34-й~Международ.~студ.~науч.~конф., Екатеринбург, 31 янв.---4 февр. 2005~г}
\def\mcnc35{Тр.~35-й~Международ.~студ.~науч.~конф., Екатеринбург, 30 янв.---3 февр. 2006~г}
\def\mcnc36{Тр.~36-й~Международ.~студ.~науч.~конф., Екатеринбург, 29 янв.---2 февр. 2007~г}
\def\mcnc37{Тр.~37-й~Международ.~студ.~науч.~конф., Екатеринбург, 28 янв.---1 февр. 2008~г}
\def\mcnc38{Тр.~38-й~Международ.~студ.~науч.~конф., Екатеринбург, 2---6 февр. 2009~г}
\def\mcnc39{Тр.~39-й~Международ.~студ.~науч.~конф., Екатеринбург, 1---5 февр. 2010~г}
\def\mcnc40{Тр.~40-й~Международ.~студ.~науч.~конф., Екатеринбург, 31 янв.---4 февр. 2011~г}
\def\tmc{Тр.~Международ.~конф.}
\def\tc{Тр.~конф.}
\def\thc{Тез.~конф.}

\def\IAUsymp{Proc.~IAU~Symp.}
\def\IAUcoll{Proc.~IAU~Colloquia}
\def\prconf{Proc.~conf.}
\def\princonf{Proc.~int.~conf.}
\def\mvrnc{Материалы~всерос.~науч.~конф}
\def\icarus{Icarus}
\def\ssr{Space~Sci.~Rev.}%
\def\aplett{Astrophys.~Lett.}%